\documentclass[conference]{IEEEtran}
\IEEEoverridecommandlockouts
\usepackage{cite}
\usepackage{amsmath,amssymb,amsfonts}
\usepackage{algorithmic}
\usepackage{graphicx}
\usepackage{textcomp}
\usepackage{xcolor}
\usepackage{url}

\usepackage{lineno}
\usepackage{listings}

\usepackage{mdwmath}
\usepackage{mdwtab}
\usepackage{multirow}
\usepackage{multicol}
\usepackage{array}
\usepackage{booktabs}

\usepackage{enumitem}
\usepackage{xspace}
\usepackage[export]{adjustbox}
\usepackage{color,soul}
\usepackage{rotating}
\usepackage{setspace}
\usepackage{amsmath} 
\usepackage{amssymb}
\usepackage{float}

\def\BibTeX{{\rm B\kern-.05em{\sc i\kern-.025em b}\kern-.08em
    T\kern-.1667em\lower.7ex\hbox{E}\kern-.125emX}}
\begin{document}

\title{Design Patterns for Blockchain-based Self-Sovereign Identity}

\author{\IEEEauthorblockN{Yue Liu}
\IEEEauthorblockA{College of Computer Science and Technology, \\China University of Petroleum (East China), China \\
yue.liu@s.upc.edu.cn}
\and
\IEEEauthorblockN{Qinghua Lu\thanks{Qinghua Lu is the corresponding author.}}
\IEEEauthorblockA{Data61, CSIRO, Australia \\
qinghua.lu@data61.csiro.au}
\and
\IEEEauthorblockN{Hye-Young Paik}
\IEEEauthorblockA{School of Computer Science and Engineering, \\University of New South Wales, Australia \\
h.paik@unsw.edu.au}
\and
\IEEEauthorblockN{Xiwei Xu}
\IEEEauthorblockA{Data61, CSIRO, Australia \\
xiwei.xu@data61.csiro.au}
}

\maketitle

\begin{abstract}
Self-sovereign identity is a new identity management paradigm that allows entities to really have the ownership of their identity data and control their use without involving any intermediary. Blockchain is an enabling technology for building self-sovereign identity systems by providing a neutral and trustable storage and computing infrastructure and can be viewed as a component of the systems. Both blockchain and self-sovereign identity are emerging technologies which could present a steep learning curve for architects. We collect and propose 12 design patterns for blockchain-based self-sovereign identity systems to help the architects understand and easily apply the concepts in system design. Based on the lifecycles of three main objects involved in self-sovereign identity, we categorise the patterns into three groups: key management patterns, decentralised identifier management patterns, and credential design patterns. The proposed patterns provide a systematic and holistic guide for architects to design the architecture of blockchain-based self-sovereign identity systems.
\end{abstract}

\begin{IEEEkeywords}
Blockchain, Self-sovereign Identity, Pattern, Architecture, Identity Management
\end{IEEEkeywords}

\section{Introduction}
\begin{figure}[t]
	\centering
	\includegraphics[width=0.7\columnwidth]{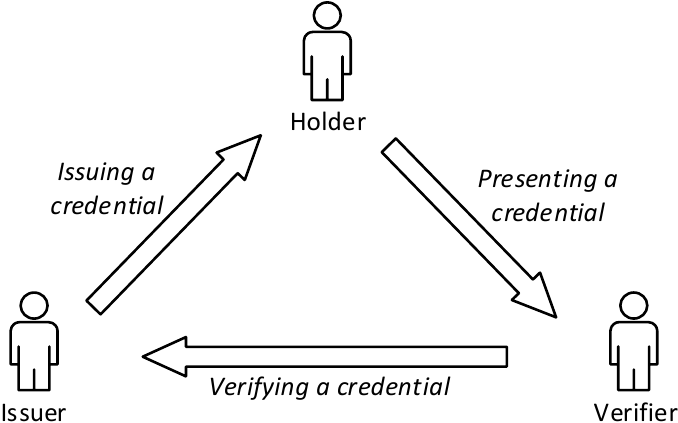}
	\caption{Identity management overview}
	\label{identityManagement}
\end{figure}

A legal entity's identity (i.e., an individual or an organisation) can be represented using a set of attributes \cite{pathToSSI} associated with the entity (such as name and address). Identity management includes maintaining the identity data and their access control. Fig.~\ref{identityManagement} depicts a conceptual overview of the main roles and their relationships in identity management. Specifically, an entity who registers an identifier in a particular system is considered as a \textit{holder} (e.g. legal individual / organisation name) of the identity data (e.g. date of birth, and role within the organisation) associated with the identifier, while all the identity data of the holder are stored with an \textit{issuer} (e.g. a government agency). Note that the holder of an identifier can sometimes also be an issuer to identify itself. A credential is a verifiable claim, which includes some facts that is attested and digitally signed by the issuer about the holder \cite{W3CCredential}. The fact in a credential can be the holder's identity data (e.g. date of birth) or other types of factual data (e.g. a GPA). After establishing a trust relationship with an issuer, anyone can be a \textit{verifier} (e.g. an employer) of a claim. A verifier requests for a specific credential (e.g. birth certificate of a person), and verifies the validity of the credential via the issuer's signature. 

Identity management is challenging if holders do not have a full control over their identity data, since the data are usually maintained at third-party issuers' sites (e.g., government agency). An issuer may disclose identity data to a third-party without the holder's knowledge. Furthermore, the issuer may be compromised, consequently resulting in identity information leakage (e.g., Aadhaar data leakage~\cite{banerjee2016aadhaar}). In addition, current credential verification processes are usually complex, costly and time consuming (e.g., taking weeks for verifying a degree). It is also possible that a significant process inconsistency occurs within current identity management systems as individuals must seek to verify (and re-verify) identities at multiple points with different service providers\footnote{\url{http://fsi.gov.au/publications/final-report/}}.

The concept of self-sovereign identity allows holders to retain ownership of their identities and control over how their identity data is used~\cite{pathToSSI}. Such a notion is increasingly popular, particularly in our digitised and privacy-sensitive society. However, for legal identities in the real world, we argue that self-sovereign identity does not mean holders can control all aspects of their identity that are maintained by the issuers. For example, holders are not able to include any additional information in their identity information (e.g., a name other than their legal name in their academic transcript), or remove existing restrictions imposed upon registration. Rather, what is achievable via self-sovereign identity is that holders can control the \textit{use of their identity data without involving any intermediary}~\cite{pathToSSI}. For example, through an effective implementation of a digital wallet for self-sovereign identity to store and manage credentials, holders who own several identifiers can choose to present any credentials associated with any of the identifiers to verifiers, without having to go through the issuers.

Blockchain is an innovative distributed ledger technology (DLT) for building new forms of decentralised software architecture, which enables agreements on transactional data sharing across a large network of untrusted participants, without relying on a central trusted authority~\cite{scheuermann2015iacr}. Identity management is considered to be one of the most innovative applications of blockchain technology~\cite{Swan:blockchain}, as blockchain can be used to build an infrastructure or ecosystem to realise self-sovereign identity, where no intermediary is needed. Many organisations (e.g., start-ups, enterprises, and governments) are currently exploring how to leverage blockchain technology to implement self-sovereign identity solutions, examples include uPort\footnote{\url{https://www.uport.me/}\label{uport}} and Hyperledger Indy\footnote{\url{https://www.hyperledger.org/projects/hyperledger-indy}}.
However, as blockchain and self-sovereign identity are both emerging technologies with limited documentation, there can be a steep learning curve for developers to design the architecture of blockchain-based self-sovereign identity systems. A recent survey by Gartner \cite{Gartner:2018:CIOSurveyBC} points to the current gap and scarcity of blockchain skills in the market.
Having a systematic and holistic guidance for the architectural design of blockchain-based self-sovereign identity systems can assist system architects and developers.
 
In this regard, this paper presents 12 design patterns for the design of blockchain-based self-sovereign identity applications. To correctly capture the use of the identities and associated credentials, we analysed the lifecycles of the three key objects in self-sovereign identity (i.e., key, identifier, and credential), and classified the patterns into three groups accordingly: key management patterns, decentralised identifier management patterns, and credential design patterns. The proposed patterns are connected to different state transitions in the lifecyles of the three key objects, which provides a guide to the architects and developers when the design patterns can be effectively used.

The remainder of this paper is organised as follows. Section~\ref{background} discusses related work. Section~\ref{patterns} presents the design patterns of blockchain-based self-sovereign identity applications with the extended pattern form in \cite{patternLanguage}. Section~\ref{conclusion} concludes the paper.

\section{Background and Related Work}
\label{background}
\subsection{Blockchain and Smart Contracts}
Blockchain is an emerging paradigm of building next generation applications in a decentralised way, with two core technologies: 1) a distributed ledger, and 2) a computing infrastructure. 

The distributed ledger maintained in a blockchain network can verify and store transactions \cite{scheuermann2015iacr}, without relying on any central trusted authority, while all participating nodes need to reach consensus on transactional data states to achieve trust. In the consensus mechanism proposed by Nakamoto, an honest majority of nodes is assumed, achieving trust without a third party intermediary through game theoretic incentives~\cite{Satoshi:bitcoin}. The data structure of blockchain is a list of identifiable blocks, and all the blocks are linked to the previous block and thus form a chain. The blocks are containers for storing transactions, which are identifiable packages carrying the changing states of data.

Blockchain provides a programmable computing infrastructure via smart contracts, which are essentially the programs deployed and run on blockchain~\cite{Omohundro:2014}. Smart contracts can express triggers, conditions to enable complex business logic. Ethereum is currently the most widely-used blockchain that supports Turing-complete smart contracts. The primary smart contract language used on Ethereum blockchain is Solidity\footnote{https://solidity.readthedocs.io/}.

Integrating blockchain technology into current software architecture brings both quality improvements and also blockchain's nature limitations. For instance, \emph{immutability} and \emph{transparency} can ensure \emph{data integrity}, while the underlying \emph{decentralisation} enhances the \emph{availability} of whole system. Nevertheless, \textit{data privacy} and \textit{scalability} are the main two limitations of public blockchains. Data privacy on public blockchain is limited because there is no privileged user, and every participant can join the network to access all the information on blockchain and validate new transactions. There are scalability limits on (i) the size of the data on blockchain, (ii) the transaction processing rate, and (iii) the latency of data transmission and commits, which is affected by the chosen consensus protocol. Furthermore, the number of transactions included in each block is also limited by the bandwidth of nodes participating in the network.

\begin{figure*}[!ht]
	\centering
	\includegraphics[width=0.76\textwidth]{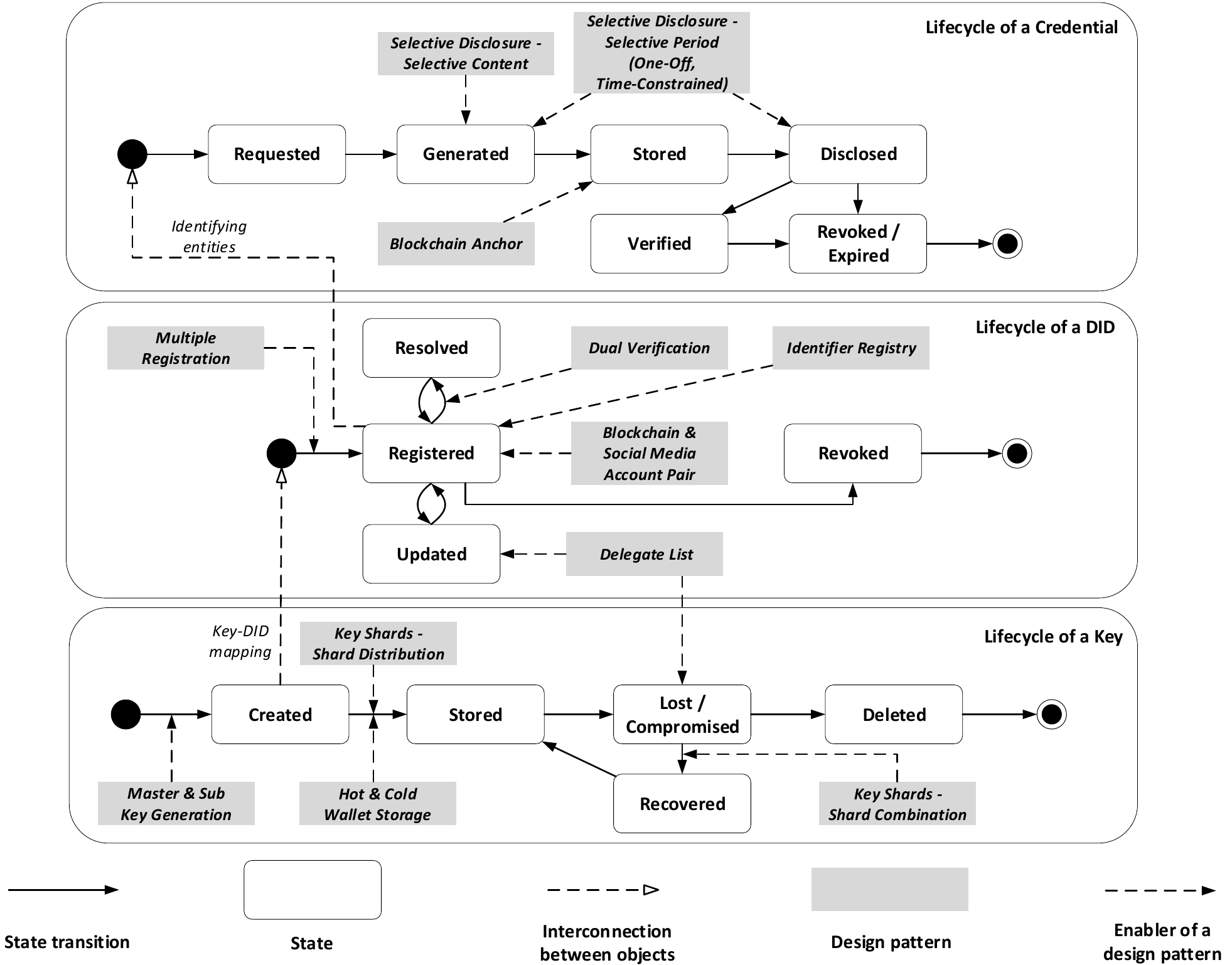}
	\caption{Lifecycles of three main objects in self-sovereign identity}
	\label{lifecycle}
\end{figure*}

\begin{table*}[htbp]
\footnotesize
\centering
\caption{Pattern collection overview}
\label{overview}
\begin{tabular}{p{0.18\columnwidth}p{0.35\columnwidth}p{1.45\columnwidth}}
\toprule

\multicolumn{1}{l}{\bf Category} &
\multicolumn{1}{c}{\bf Name} &
\multicolumn{1}{c}{\bf Summary}\\
\midrule

\multirow{6}{0.18\columnwidth}{\textbf{Key Management Patterns}} & \multirow{2}{0.32\columnwidth}{Master \& Sub Key} &  Each entity has a master-key for managing sub-keys which are used for signing transactions for different identity accounts.\\
\cmidrule(l){2-3}
& \multirow{2}{0.32\columnwidth}{Hot \& Cold Wallet Storage} & \multirow{2}{1.45\columnwidth}{An entity can maintain a hot wallet connecting to internet and a cold wallet which is kept offline.}\\ \\
\cmidrule(l){2-3}
& \multirow{1}{0.32\columnwidth}{Key Sharding} & A key can be split up into several different pieces, and restored using enough key pieces.\\
\cmidrule(l){1-3}
\multirow{9}{0.18\columnwidth}{\textbf{DID Management Patterns}} & \multirow{1}{0.32\columnwidth}{Identifier Registry} & The identifier registry maintains bindings between an identifier and the address of an identity attribute. \\
\cmidrule(l){2-3}
&\multirow{1}{0.32\columnwidth}{Multiple Registration} & Each entity can register a unique identifier for each relationship. \\
\cmidrule(l){2-3}
& \multirow{2}{0.32\columnwidth}{Bound with Social Media} & \multirow{2}{1.45\columnwidth}{A bidirectional binding is established between social media profile and blockchain-based identity.} \\ \\
\cmidrule(l){2-3}
& \multirow{2}{0.32\columnwidth}{Dual Resolution} & To establish interactions, two entities can mutually acquire each other's DDO for verification and communication information.\\
\cmidrule(l){2-3}
& \multirow{1}{0.32\columnwidth}{Update by Delegates} & Each identifier maintains a list of delegates that can help the user recover the identity.\\
\cmidrule(l){1-3}
\multirow{9}{0.18\columnwidth}{\textbf{Credential Design Patterns}} & \multirow{1}{0.35\columnwidth}{Selective Content Generation} & An issuer generates a customised credential according to a holder’s specific requirements about credential contents. \\
\cmidrule(l){2-3}
& \multirow{2}{0.32\columnwidth}{Time-Constrained Access} & \multirow{2}{1.45\columnwidth}{ A holder can share a link which is redirected to the credential content only for a specified period of time.} \\ \\  
\cmidrule(l){2-3}
& \multirow{2}{0.32\columnwidth}{One-Off Access} & A holder can share a one-off link which is redirected to the credential content one time only to satisfy a temporary identification request.\\
\cmidrule(l){2-3}
& \multirow{2}{0.32\columnwidth}{Anchoring to Blockchain} & \multirow{2}{1.45\columnwidth}{ Periodically sending the unique hash value of off-chain data to blockchain.} \\ \\
\bottomrule
\end{tabular}
\end{table*}

\subsection{Self-Sovereign Identity}
Identity management is a fundamental requirement in our digitised society, since every entity is likely to have multiple identities for different people or organisations \cite{modinis2005common,dabrowski2008generic}. However, Internet users do not generally have complete control over their digital identities stored by third-party issuers (e.g. social networking sites or employers), which may disclose one's identity information to others without their permission and/or knowledge. 

In self-sovereign identity \cite{pathToSSI}, identity owners are central to the administration of the identities, in the sense that they are able to manage their identities on personal mobile devices or cloud, and interact with multiple service providers. To implement self-sovereign identity, the W3C Community Group specifies a standard for decentralised identifier (DID) \cite{W3CDID}, which contains human-readable information and can be used across different platforms. 
A DID is a URL that refers to an entity for trusted interactions. Each DID bonds to a DID document (DDO) which describes how to use the specific DID through some given properties, such as \emph{context}, \emph{id}, \emph{public key}, and \emph{service endpoint}. 
\emph{context} expresses the system environment for information exchange between two DIDs; \emph{id} is the DID described by this DDO; \emph{publicKey} is for digital signatures and cryptographic operations; and \emph{service} denotes service endpoints used for interactions among DIDs. 
A DID and its corresponding DDO do not contain any identity data which are stored in personal devices or database owned by the identity service providers. 
A credential \cite{W3CCredential} is a verifiable claim, carrying particular identity information attributes that an issuer provides to attest to some specific characteristics of an entity. Within self-sovereign identity, entities are able to issue digitally-signed credentials about themselves and others linked to their DIDs, and these credentials can be verified by anyone else. Blockchain has been widely recognised as a viable technology for enabling DID in terms of data integrity and privacy. As self-sovereign identity needs no intermediary, it aligns with the design nature of blockchain which eliminates the need for a centralised authority.

\subsection{Related Work}
Previous works have characterised and applied design patterns for smart contract and blockchain applications \cite{xu2018pattern, smartContractICBC}. originChain \cite{IEEESoftware2017, originChain} adopts design patterns for data management and smart contract design to improve the system adaptability. uBaaS encapsulates design patterns as services to facilitate the development of blockchain-based applications \cite{ubaas}. Bartoletti and Pompianu \cite{bartoletti2017empirical} conduct an empirical analysis of smart contracts, in which they collected hundreds of smart contracts and divided them into nine categories: token, authorisation, oracle, randomness, poll, time constraint, termination, math and fork check. Eberhardt and Tai \cite{eberhardt2017or} propose five patterns, including challenge response pattern, off-chain signatures pattern, content-addressable storage pattern, delegated computation pattern, and low contract footprint pattern, mainly focusing on the separation of on-chain and off-chain for data and computation. Zhang et al. \cite{zhang2017applying} share the experience of designing a blockchain-based healthcare platform, to which they apply four object-oriented software patterns to improve the application scalability. Wohrer and Zdun \cite{wohrer2018smart} collect six design patterns to address security issues of smart contract design.

There have been significant efforts in both industry and academia to address many of the identity management challenges using blockchain technology~\cite{muhle2018survey,lim2018blockchain}. Many organisations and companies have been developing self-sovereign identity platforms using blockchain, including Sovrin (Hyperledger Indy)~\cite{Sovrin}
, uPort (Ethereum)~\cite{lundkvist2017uport}, and Blockstack (Namecoin)~\cite{ali2016blockstack}. Decentralised Identity Foundation\footnote{\url{https://identity.foundation/}} aims to develop an open ecosystem for decentralised identity, and the Internet Identity Workshop\footnote{\url{https://internetidentityworkshop.com/}} seeks to find, investigate, and solve identity issues. There are also projects designing and developing self-sovereign identity systems~\cite{grather2018blockchain,liang2017integrating,indyKYC,stokkink2018deployment,takemiya2018sora}.

Compared with the existing works, our study focuses on the design patterns of blockchain-based self-sovereign identity, aiming to facilitate the design and development of self-sovereign identity applications. Some collected patterns are already applied in the related works, for instance, \emph{Identifier Registry}, \emph{Multiple Registration}, and \emph{Selective Content Generation} can be found in the above projects and studies.

\begin{figure*}[!ht]
	\centering
	\includegraphics[width=0.7\textwidth]{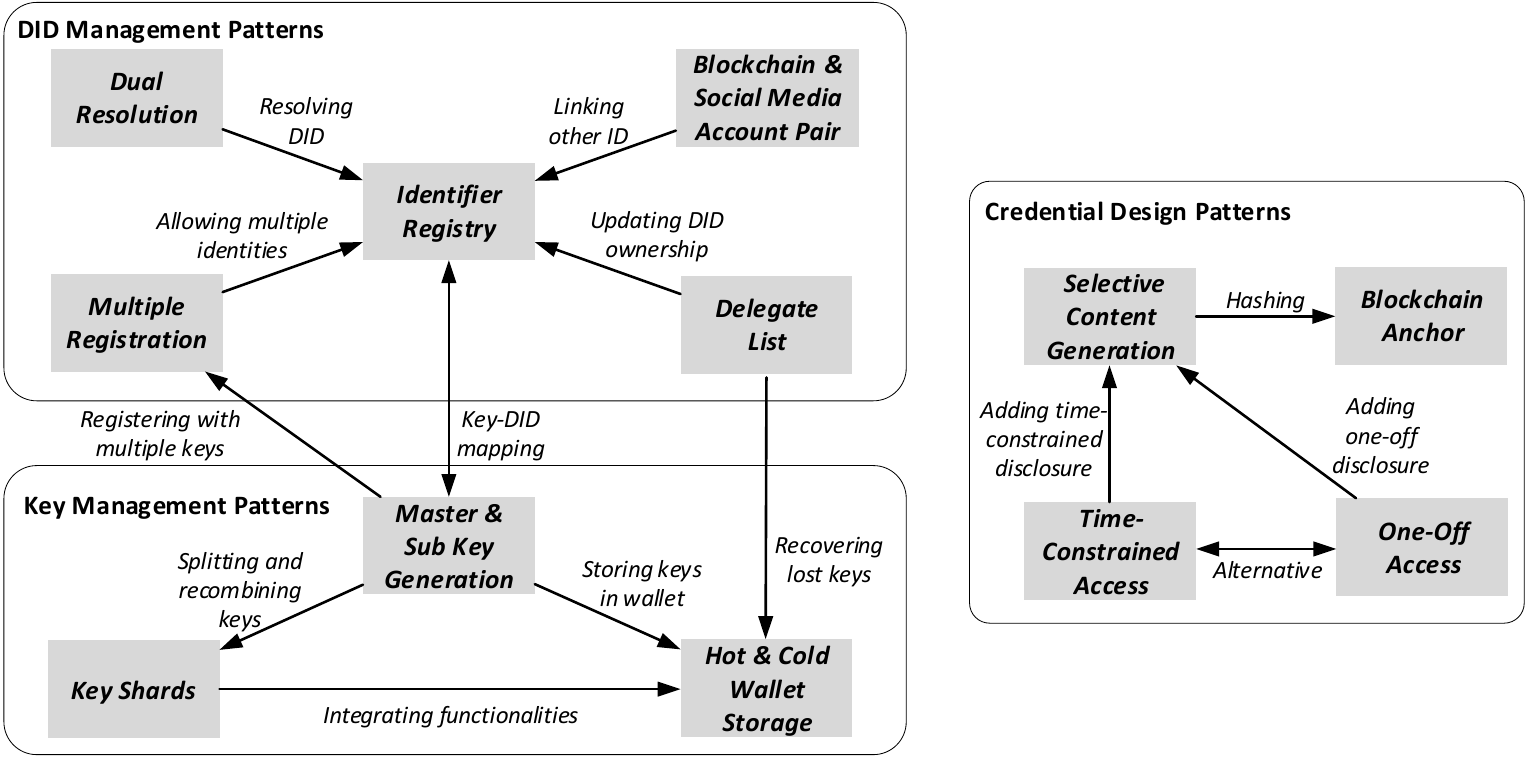}
	\caption{Pattern relation overview}
	\label{relationshipPic}
\end{figure*}

\section{Patterns for Blockchain-based Self-Sovereign Identity}
\label{patterns}

This section introduces 12 patterns for the design of blockchain-based self-sovereign identity applications. We present the patterns in three groups: key management, DID management and credential design. The patterns aim to  make better use of three main objects in self-sovereign identity -- key, DID, and identity credential,  by understanding their use in different stages of the lifecycles.

Fig.~\ref{lifecycle} and Table~\ref{overview} offer an overview of these patterns, while Fig. \ref{relationshipPic} gives a glimpse of the relations between the patterns collected in this study. We describe each pattern using the extended pattern form in \cite{patternLanguage}. This includes the pattern name, a short summary, the usage context, the problem statement, a discussion on the forces leading to the problem difficulty, the solution and its consequences, and some examples of real-world known uses of the pattern. Note that the forces are identified with the quality attributes which may affect the application, and the solution sometimes proposes a trade-off to mitigate the dilemma.

\subsection{Key Management Patterns} 

\subsubsection{Master \& Sub Key Generation} \noindent \par
\label{master-sub}

\begin{figure}[!ht]
	\centering
	\includegraphics[width=0.47\columnwidth]{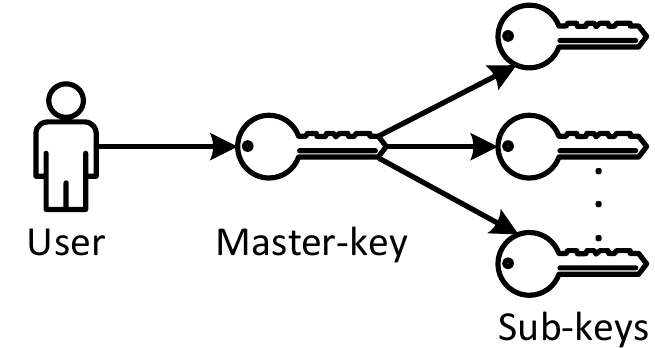}
	\caption{Master \& Sub Key Generation Pattern}
	\label{master-subPic}
\end{figure}

\vspace{0.5em}\noindent \textbf{Summary:} 
Each entity has a master-key for managing sub-keys which are used to sign transactions for different identity accounts. Fig.~\ref{master-subPic} is a graphical representation of the pattern.

\vspace{0.5em}\noindent \textbf{Context:} 
Public key cryptography and digital signatures are used to identify accounts and authorise transactions submitted to a blockchain. 

\vspace{0.5em}\noindent \textbf{Problem:} 
Using a single key for all transactions has serious privacy implication for an identity owner since transactions can be correlated to expose all the identities an entity holds.

\vspace{0.5em}\noindent \textbf{Forces:} The problem requires 
balancing the following forces:
\begin{itemize}
  \item \textit{Identifiability.} An entity needs to create an identity for sending transactions in the blockchain network.
  
  \item \textit{Re-identification.} Each entity can have more than one identities in the real world. For example, one person is both a client of a bank and a patient of a hospital. If an entity handles all the identities via a single key, its privacy may be violated as all the transactions on blockchain are transparent and can be correlated.
  
  \item \textit{Security.} There is no a standard approach to protect or recover users' secret keys.
\end{itemize}

\vspace{0.5em}\noindent \textbf{Solution:} 
Each entity can have a master-key to manage sub-keys which are used for signing messages under different identities. For example, a person can have a sub-key for the student identity and another sub-key for the company intern identity. Each sub-key is linked to a unique identifier and stored as part of the identifier's data in the identifier registry, which can be updated using the master-key. The use of master-key must be minimised (i.e., only used for controlling sub-keys) due to its importance.

\vspace{0.5em}\noindent \textbf{Consequences:} 

Benefits:
\begin{itemize}
  \item \textit{Identifiability.} Identifiability is preserved as each transaction is signed by a private key.
  
  \item \textit{Privacy.} Each sub-key maintains its own identity.
  The transactions sent under different identities are less likely to be mapped and correlated
  in order to re-identify the user.
  
  \item \textit{Availability.} If a sub-key is lost or compromised, the master-key can be used to revoke the lost or compromised key and update it with a new sub-key.
\end{itemize}

Drawbacks: 
\begin{itemize}
   \item \textit{Security.} If the master-key is lost or compromised, the identity owner loses control of all the sub-keys which means loss of control of all the owned identities. 
\end{itemize}

\vspace{0.5em}\noindent \textbf{Related patterns:} 
\begin{itemize}
    \item \emph{Hot \& Cold Wallet Storage} (Section~\ref{hot-cold}) Entities' keys are stored in wallet applications.

    \item \textit{Key Shards} (Section~\ref{Key Shards}) Keys can be protected and recovered by \textit{Key Shards}.
    
    \item \emph{Identifier Registry} (Section~\ref{identifier registry}) Entities can use sub-keys to register identifiers in \emph{Identifier Registry}.

    \item \emph{Multiple Registration} (Section \ref{multiple identifier}) Sub-keys are used to register identifiers for different identity relationships.
\end{itemize}

\vspace{0.5em}\noindent \textbf{Known uses:}
\begin{itemize}
  \item \emph{uPort}\textsuperscript{\ref{uport}} A uPort user interacts with application smart contracts via  a proxy smart contract, thus the public key of the proxy contract is considered as a layer of indirection  between the user’s private key and the target application contract.
  
  \item \emph{Ethereum}\footnote{\url{https://ethereum.org/}\label{ethereum}} Implementing the Ethereum ERC725 standard key management functions\footnote{\url{https://github.com/ethereum/EIPs/issues/725}} requires the deployment of ERC 725 identity smart contracts, which act as unique identifiable proxy accounts and used by humans, groups, organisations, objects and machines.
  
  \item \emph{Trinity}\footnote{\url{https://trinity.iota.org/}} Trinity is a wallet application of IOTA ledger\footnote{\url{https://www.iota.org/}}. Each account has a seed acting as a master-key, to control the addresses and IOTA tokens of that specific account.
\end{itemize}

\subsubsection{Hot \& Cold Wallet Storage}  \noindent \par
\label{hot-cold}

\begin{figure}[!ht]
	\centering
	\includegraphics[width=0.418\columnwidth]{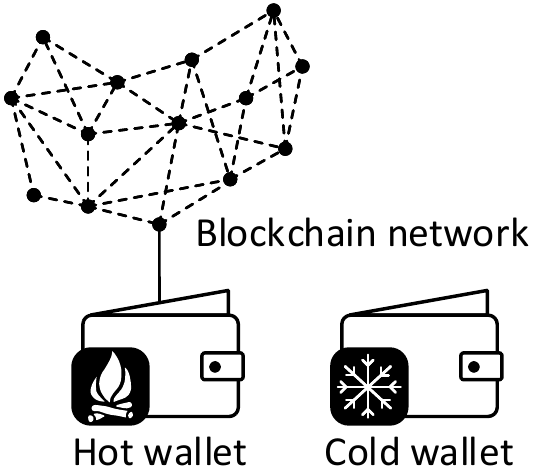}
	\caption{Hot \& Cold Wallet Storage Pattern}
	\label{hot-coldPic}
\end{figure}

\vspace{0.5em}\noindent \textbf{Summary:} An entity can maintain a hot wallet connecting to internet and a cold wallet which is kept offline. Fig. \ref{hot-coldPic} is a graphical representation of the pattern.

\vspace{0.5em}\noindent \textbf{Context:} As a blockchain network participant, one entity can rely on so-called ``wallets" to manage its accounts and interact with blockchain.

\vspace{0.5em}\noindent \textbf{Problem:} An entity's wallet may suffer malicious attacks, leading to key theft. The attacker can send transactions under that entity's name to blockchain using a compromised key.

\vspace{0.5em}\noindent \textbf{Forces:} The problem requires balancing the following forces:
\begin{itemize}
  \item \textit{Cyber security. A key may be hacked when being stored in a device connected to internet.}
  
  \item \textit{Usability.} Some keys may be frequently used by blockchain participants while other keys might act as backup. 
\end{itemize}

\vspace{0.5em}\noindent \textbf{Solution:} Users can choose to store keys in two types of wallets, namely hot wallet and cold wallet. Hot wallet refers to the blockchain gateways that are connected to Internet. Through a hot wallet, a user is able to directly conduct specific operations (e.g. generation) to its accounts and related decentralised identifiers stored on-chain. Cold wallet refers to key storage that is off-line, keeping the accounts from being hacked. A cold wallet can be any device disconnected from the internet or even a paper recording an entity's keys. When the keys stored in a cold wallet are required for signing transactions, the user needs to connect the cold wallet device to a computer and copy-paste the key in the relevant field. A user can combine these two wallets: storing accounts that are frequently used in a hot wallet while using a cold wallet to keep those that are not used often.

\vspace{0.5em}\noindent \textbf{Consequences:}

Benefits:
\begin{itemize}
  \item \textit{Secure storage.} Cold wallets are separated from internet, which provide secure storage for entities' keys.
  
  \item \textit{Usability.} Such a secure storage also preserves the usability of keys, as once a cold wallet is connected to Internet, an entity can utilize those keys.
\end{itemize}

Drawbacks: 
\begin{itemize}
   \item \textit{Security.} Hot wallets store one's secret keys on-line, which is still vulnerable to theft. 
   
   \item \textit{Usability.} Cold wallets are more secure than hot wallets but less convenient to use, as the user has to connect to the cold wallet which might not be around.
\end{itemize}

\vspace{0.5em}\noindent \textbf{Related patterns:}
\begin{itemize}
     \item \emph{Master \& Sub Key Generation} (Section \ref{master-sub}) Wallet applications are utilised to stored users' keys.

    \item \emph{Key Shards} (Section~\ref{Key Shards}) Splitting and recombining a key should be conducted in a wallet application.

     \item \textit{Delegate List} 
     (Section~\ref{recovery delegate registry}) 
     When being integrated into wallet applications, predefined delegates can replace key ownership if a key is compromised.
\end{itemize}

\vspace{0.5em}\noindent \textbf{Known uses:}
\begin{itemize}
   \item \textit{MyEtherWallet}\footnote{\url{https://www.myetherwallet.com/}} Ethereum blockchain network offers a software , MyEtherWallet, as hot wallet to users for instant payment and withdrawal. With a visual interface, it is easy for users to operate without inputting complicated commands.
   
   \item \textit{Trezor}\footnote{\url{https://trezor.io/}} Trezor is a cryptocurrency hardware wallet, designed to store and encrypt users' coins, passwords and other digital keys. It is a single-purpose computer with independent memory to save all private data.
   
   \item \textit{Ledger}\footnote{\url{https://www.ledger.com/}} Ledger provides hardware wallet products to stores users' private keys in a secure hardware device, protecting the cryptocurrencies. 
\end{itemize}

\subsubsection{Key Shards}  \noindent \par
\label{Key Shards}

\begin{figure}[t]
	\centering
	\includegraphics[width=0.65\columnwidth]{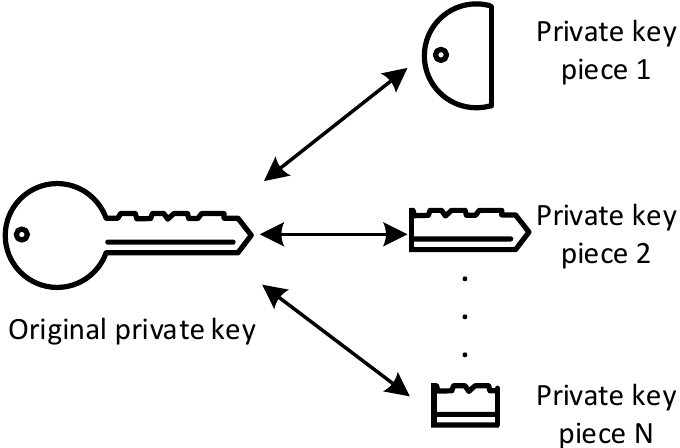}
	\caption{Key Shards Pattern}
	\label{keyShardingPic}
\end{figure}

\vspace{0.5em}\noindent \textbf{Summary:} A key can be split up into several different pieces, and restored using enough key pieces. Fig.~\ref{keyShardingPic} is a graphical representation of the pattern.

\vspace{0.5em}\noindent \textbf{Context:} In self-sovereign identity, a participant may have multiple keys, for instance, signing key for transaction authorisation, public/private key pair for encryption/decryption, etc. Consequently, key management is significant to the users, especially credential issuers and holders.

\vspace{0.5em}\noindent \textbf{Problem:}  A user may lose or forget his/her secret keys under some circumstances, e.g. the device containing the keys is lost or broken. Losing the keys denotes that the owner could lose control over its blockchain accounts in self-sovereign identity and the related identities.

\vspace{0.5em}\noindent \textbf{Forces:} 
\begin{itemize}
  \item \textit{Loss of keys.} Private keys are usually long and hard to remember. Also, many blockchain platforms do not provide a sound mechanism to recover the signing keys. 
  
  \item \textit{Centralisation.} A blockchain user's keys are usually stored in a wallet application installed on a mobile device, and such a centralised approach to maintain keys may cause a single-point of failure. Once the device is lost or hacked, the user may lose the control of all keys.
\end{itemize}

\vspace{0.5em}\noindent \textbf{Solution:} To protect the security of a secret key, one can spilt that key into several pieces as its requirement, and define a regrouping threshold. The key pieces can be kept in any way the user prefers, e.g., wrote on a paper and locked in a safe box, given to family and friends, etc. When a key is lost, the user needs to regain enough key pieces (more than the preset regrouping threshold), and these pieces can help rebuild the complete key.

\vspace{0.5em}\noindent \textbf{Consequences:} 

Benefits:
\begin{itemize}
  \item \textit{Tolerance of lost key.} A lost key can be recovered by its shards, which provides great convenience to blockchain users. 
  \item \textit{Decentralisation.} The shards are stored in a decentralised way, which reduces the risk of losing all shards in an attack.
  \item \textit{Flexibility.} An entity does not have to input all but just enough key shard when recombining a lost key.
\end{itemize}

Drawbacks: 
\begin{itemize}
   \item \textit{Cost.} Maintaining key shards needs extra vigor. If a key shard is lost, there is no way to restore it.
   \item \textit{Security.} Having multiple key shards provides multiple targets to attack.
\end{itemize}

\vspace{0.5em}\noindent \textbf{Related patterns:} 
\begin{itemize}
\item \textit{Master \& Sub Key Generation} (Section~\ref{master-sub}) \emph{Key Shards} is applied to split and recombine entities' keys.

\item \textit{Hot \& Cold Wallet Storage} (Section~\ref{hot-cold}) The key splitting and recombining functionalities should be integrated into wallet applications.

\item \textit{Delegate List (Section}~\ref{recovery delegate registry}) \emph{Key Shards} can restore a lost key, while \emph{Delegate List} aims to replace a compromised key with a new one.
\end{itemize}

\vspace{0.5em}\noindent \textbf{Known uses:}
\begin{itemize}
 \item \textit{Parity}\footnote{\url{https://www.parity.io/}} For each private key of an account, Parity distributes 12-word phrase acting as an additional backup. If a user loses the private key, this phrase can help fully recover it.
  
  \item \textit{Crypto++}\footnote{\url{https://cryptopp.com/}} A free and open-source C++ class library of cryptographic algorithms and schemes, which implements the Shamir's Secret Sharing scheme: spliting up a secret into defined number of pieces, and restoring the original secret when given enough secret pieces.
\end{itemize}

\subsection{DID Management Patterns}

\subsubsection{Identifier Registry}  \noindent \par
\label{identifier registry}

\begin{figure}[!ht]
	\centering
	\includegraphics[width=0.45\columnwidth]{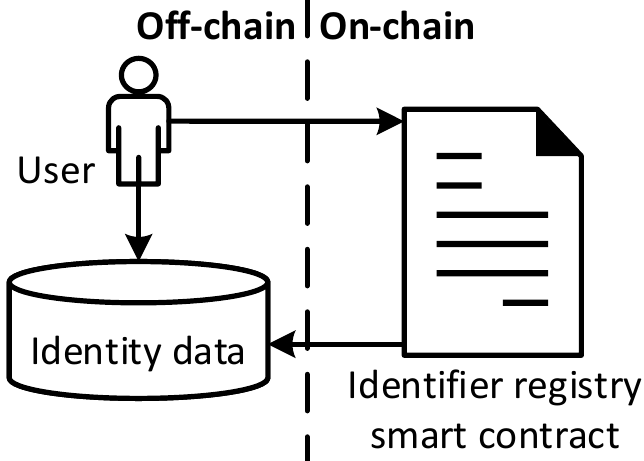}
	\caption{Identifier Registry Pattern}
	\label{identifierRegistryPic}
\end{figure}

\vspace{0.5em}\noindent \textbf{Summary:} The identifier registry maintains bindings between an identifier and the address of an identity attribute (e.g. name, profile, picture). Fig.~\ref{identifierRegistryPic} is a graphical representation of the pattern.

\vspace{0.5em}\noindent \textbf{Context:} 
Identity is defined as sets of attributes related to an entity\footnote{\url{https://www.iso.org/standard/77582.html}}. 
In software applications, identity attribute data needs to be accessed for a specific purpose. An identifier is a globally unique persistent series of digits and/or characters that is used to uniquely identify an entity (e.g. human, organisation, device) within one domain and can be used to retrieve the storage location of the identity attribute data. A Decentralised Identifier (DID) is a new type of identifier which is designed for cryptographically verifiable self-sovereign identity.

\vspace{0.5em}\noindent \textbf{Problem:} 
In traditional centralised software systems, mappings between an identifier and the identity data storage location is maintained by a centralised single authority which may become a potential single point of failure.

\vspace{0.5em}\noindent \textbf{Forces:} The problem requires 
balancing the following forces:
\begin{itemize}
  \item \textit{Upgradability.} The need to upgrade identity data over time is ultimately necessary for software applications.
  \item \textit{Scalability.} Blockchain provides limited scalability because data is replicated across all nodes and kept permanently.
  \item \textit{Cost.} Storing data to blockchain may have monetary cost (if choosing a public blockchain), and also occupies the physical storage of all participating nodes.
\end{itemize}

\vspace{0.5em}\noindent \textbf{Solution:} 
Implementing an identifier registry designed as a smart contract to maintain bindings between an identifier and the location of associated off-chain identity data attributes. This identifier registry smart contract is the main entry point for accessing the attributes of an identity, which can map each identifier to a storage (e.g. IPFS, Dropbox, etc.) location for the respective identity attributes (e.g. an IPFS hash linking to the IPFS storage location containing the user's identity attributes). Only the identifier owner is allowed to update the storage location of identity attributes. Each identifier points to an identifier document which describes how to use that specific identifier, e.g. public keys used for digital signatures, service endpoints for interaction.

\vspace{0.5em}\noindent \textbf{Consequences:} 

Benefits:
\begin{itemize}
  \item \textit{Upgradability.} Through the cryptographic access controls built into the blockchain, it is guaranteed that only the owner of the identity (i.e. the key owner) has the right to modify the data reference in identifier registry.
  \item \textit{Scalability.} The data structure of identifier registry is simple and lightweight, which only stores identifiers and locations of identity attributes. 
  \item \textit{Cost.} If a public blockchain is used, the cost is low since the data size of identifiers and identity storage locations is fixed.
\end{itemize}

Drawbacks: 
\begin{itemize}
   \item \textit{Integrity.} The off-chain identity data store might not be as secure as blockchain. The raw data may be changed without authorisation. If IPFS is used for identity data storage, the change will be detected. However, without additional measures, it will neither be possible to recover the original data nor to prevent the change from happening in the first place.
   \item \textit{Data loss.} Since the raw data is stored off-chain, it may be deleted or lost. 
\end{itemize}

\vspace{0.5em}\noindent \textbf{Related patterns:} 
\begin{itemize}
\item \textit{Multiple Registration} (Section~\ref{multiple identifier}) An entity can register multiple identifiers at once in \emph{Identifier Registry}.

\item \textit{Blockchain \& Social Media Account Pair} (Section~\ref{social media}) \emph{Identifier Registry} maintains the mapping between an entity's on-chain identity and social media.

\item \textit{Dual Resolution} (Section~\ref{Dual Resolution}) Entities resolve each other's identifiers in \emph{Identifier Registry}.

\item \textit{Delegate List} (Section~\ref{recovery delegate registry}) Predefined delegates can change the binding key of a compromised identifier.

\item \textit{Master \& Sub Key Generation} (Section~\ref{master-sub}) Entities register identifiers in \emph{Identifier Registry} through their sub-keys.

\item \textit{Content-Addressable Storage Pattern}~\cite{eberhardt2017or} \emph{Identifier Registry} is similar to \textit{Content-Addressable Storage Pattern}, storing raw data off-chain while publishing a reference in smart contract.

\item \textit{Flyweight Pattern}~\cite{zhang2017applying} \emph{Identifier Registry} and \textit{Flyweight Pattern} both include a registry to maintain the mapping of identifiers and the respective reference pointing to off-chain data.
\end{itemize}

\vspace{0.5em}\noindent \textbf{Known uses:}
\begin{itemize}
  \item \emph{uPort}\textsuperscript{\ref{uport}} The registry contract maintains mappings between a uPort identifier and an IPFS hash linking to a data structure storing an entity's attributes.
  
  \item \emph{Sovrin}\footnote{\url{https://sovrin.org/}\label{sovrin}} Sovrin offers a registry for decentralised identifiers and the associated public keys and communications endpoints. The operations (i.e. registration, update, resolution, and revocation) are all determined by the Sovrin protocol.
  
  \item \emph{Jolocom}\footnote{\url{https://jolocom.io/}\label{jolocom}}  Jolocom distributed identity system integrates Ethereum blockchain and Interplanetary File System (IPFS). A deployed smart contract maintains the mapping from a DID to the corresponding DDO stored in IPFS.
\end{itemize}

\subsubsection{Multiple Registration}  \noindent \par
\label{multiple identifier}

\begin{figure}[!ht]
	\centering
	\includegraphics[width=0.63\columnwidth]{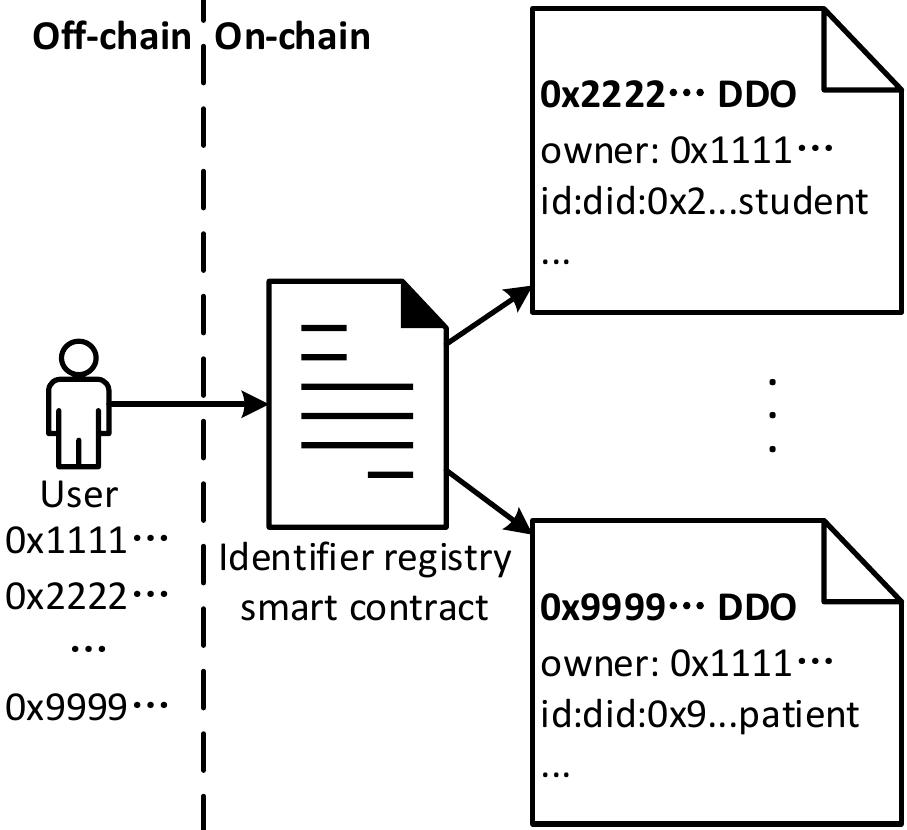}
	\caption{Multiple Registration Pattern}
	\label{multipleRegistrationPic}
\end{figure}

\vspace{0.5em}\noindent \textbf{Summary:} 
Each entity can register a unique identifier for every relationship (i.e. every identity) they have. Fig.~\ref{multipleRegistrationPic} is a graphical representation of the pattern.

\vspace{0.5em}\noindent \textbf{Context:} 
An identifier is used to uniquely identify an entity and to retrieve the identity attribute data. 

\vspace{0.5em}\noindent \textbf{Problem:} 
Sending all transactions using a single identifier has serious privacy implication for an entity since these transactions can be correlated to expose all the identities this entity holds.

\vspace{0.5em}\noindent \textbf{Forces:} The problem requires 
balancing the following forces:
\begin{itemize}
  \item \textit{Transparency.} All the historical transactions on blockchain can be accessed by every participant within the same blockchain network. The transactions on a public blockchain are also accessible to everyone on the internet.
  
  \item \textit{Security.} There is no standard approach to protect or recover users' secret keys.
\end{itemize}

\vspace{0.5em}\noindent \textbf{Solution:} 
Each entity can establish a unique identifier for every relationship (i.e. every identity) they have, which allows keeping interactions with one entity entirely separate from any other entity. For example, the relationship a person builds with a hospital is completely separate to the one that is established with a university. Neither the hospital nor the university could proactively use the identifiers to correlate this person's activities.

\vspace{0.5em}\noindent \textbf{Consequences:}
Benefits:
\begin{itemize}
  \item \textit{Privacy.} This pattern tackles with the nature transparency of blockchain to some extent, as an entity's activities in different identity relationships can hardly be correlated.
  
  \item \textit{Availability.} The loss of signing key of one identifier does not affect the other identifiers.
\end{itemize}

Drawbacks: 
\begin{itemize}
   \item \textit{Cost.} Compared to a single identifier, multiple identifiers cost more for identifier registration and updates.
\end{itemize}

\vspace{0.5em}\noindent \textbf{Related patterns:} 
\begin{itemize}
\item \textit{Identifier Registry} (Section~\ref{identifier registry}) Registered identifiers are stored in \textit{Identifier Registry}.

\item \textit{Master \& Sub Key Generation} (Section~\ref{master-sub}) Entities use different sub-keys to conduct \emph{Multiple Registration}.
\end{itemize}

\vspace{0.5em}\noindent \textbf{Known uses:}
\begin{itemize}
  \item \emph{Sovrin}\textsuperscript{\ref{sovrin}} Sovrin suggests users to use a separate identifier for every relationship. Consequently, if a relationship suffers a breach and the identifier is compromised, the user can still have normal interactions in other relationships.
  
  \item \textit{Blockstack}\footnote{\url{https://blockstack.org/}\label{blockstack}} Blockstack allows entities to create as many identities as they want. Each identity is represented by an identifier and a secret key. Entities can utilise their identities to sign in different decentralised applications.
  
  \item \textit{DAML}\footnote{\url{https://daml.com/}\label{DAML}} In DAML ledger, participant nodes can use human-readable strings as identifiers to identify themselves. A real-world entity is able to possess multiple identifiers in the same ledger network to denote different identities.
\end{itemize}

\subsubsection{Blockchain \& Social Media Account Pair}  \noindent \par
\label{social media}

\begin{figure}[!ht]
	\centering
	\includegraphics[width=0.64\columnwidth]{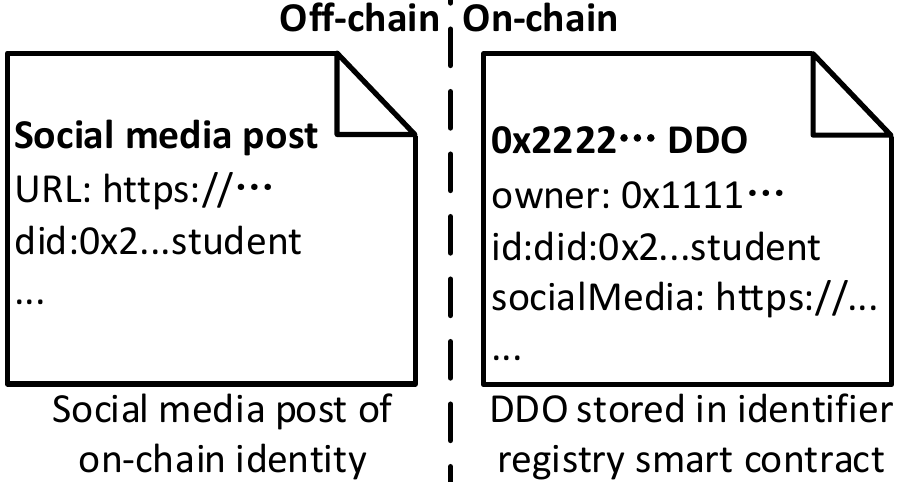}
	\caption{Blockchain \& Social Media Account Pair Pattern}
	\label{socialMediaPic}
\end{figure}

\vspace{0.5em}\noindent  \textbf{Summary:}
A bidirectional binding is established between social media profile and blockchain-based identity. Fig.~\ref{socialMediaPic} is a graphical representation of the pattern.

\vspace{0.5em}\noindent \textbf{Context:} 
Social media profiles can be considered as one of the most important assets, which are critical to achieve more exposure on the internet, attract more attention, or improve online reputation. The trustworthiness of a social media profile can be improved by verifying the account using traditional identity issued by some central authority. On the other hand, blockchain provides a decentralised infrastructure for self-sovereign identity, where entities are in control over their own identities. To ensure the trustworthiness, some identity blockchains (e.g. Hyperledger Indy) are designed as public permissioned blockchains, which are governed by a group of participants.

\vspace{0.5em}\noindent \textbf{Problem:} 
In addition to verification by some certain people or central authorities, a user can link his/her social media profile (e.g. Twitter) to his/her identity registered on blockchain to improve the trustworthiness of both social media profile and blockchain-based identity. The problem here is how to bind a social media profile with the corresponding blockchain-based identity to ensure mapping.

\vspace{0.5em}\noindent \textbf{Forces:} The problem requires 
balancing the following forces:
\begin{itemize}
  \item \textit{Authoritative source.} A 1-to-1 mapping is required between a social media account and its corresponding blockchain-based identity.
  
  \item \textit{Security.} The mapping needs to be stored securely from tampering.
  
  \item \textit{Verified accounts.} Social media applications verify accounts via traditional identity documents. 
\end{itemize}

\vspace{0.5em}\noindent \textbf{Solution:} 
An entity can create an attribute of social media in the identifier document. Signing the attribute with the blockchain signing key creates a claim that the blockchain-based identity controls the social media account. The attribute also contains a URL which links to a social media post stating that the social media account also controls this particular blockchain identity. Thus, a two-way link is established for connecting the blockchain identity with the social media profile. The two directional binding makes sure that that the social media profile and blockchain-based identity have a 1-to-1 mapping.

\vspace{0.5em}\noindent \textbf{Consequences:} 

Benefits:
\begin{itemize}
  \item \textit{Authoritative source.} The trustworthiness of social media account and blockchain-based identity can both be improved by binding them together.
  
  \item \textit{Secure storage.} Blockchain provides a secure data store through distributed ledger technology.
  
  \item \textit{Verified accounts.} A social media verifies the legal identity of a user.
\end{itemize}

Drawbacks: 
\begin{itemize}
   \item \textit{Trustworthiness.} The trustworthiness of social media account relies on the verification process while the the trustworthiness of blockchain-based identity depends on the trustworthiness of trusted participants.  
\end{itemize}

\vspace{0.5em}\noindent \textbf{Related patterns:} 
\begin{itemize}
\item \textit{Identifier Registry} (Section~\ref{identifier registry}) \textit{Identifier Registry} maintains the mapping of an entity's on-chain identity and social media.
\end{itemize}

\vspace{0.5em}\noindent \textbf{Known uses:}
\begin{itemize}
  \item \emph{Onename}\footnote{\url{https://www.onename.com/}} Onename is a registrar for Blockstack which is a decentralised naming and storage system. With Onename, a user can easily register a blockchain ID and create corresponding profile on Blockstack, and completely has control of his/her blockchain ID and profile. Users can can link their Blockstack profiles with existing online social media accounts (Twitter, Facebook, Github, etc), and also embed their blockchain ID on their social media posts.
\end{itemize}

\subsubsection{Dual Resolution} \noindent \par
\label{Dual Resolution}

\begin{figure}[!ht]
	\centering
	\includegraphics[width=0.63\columnwidth]{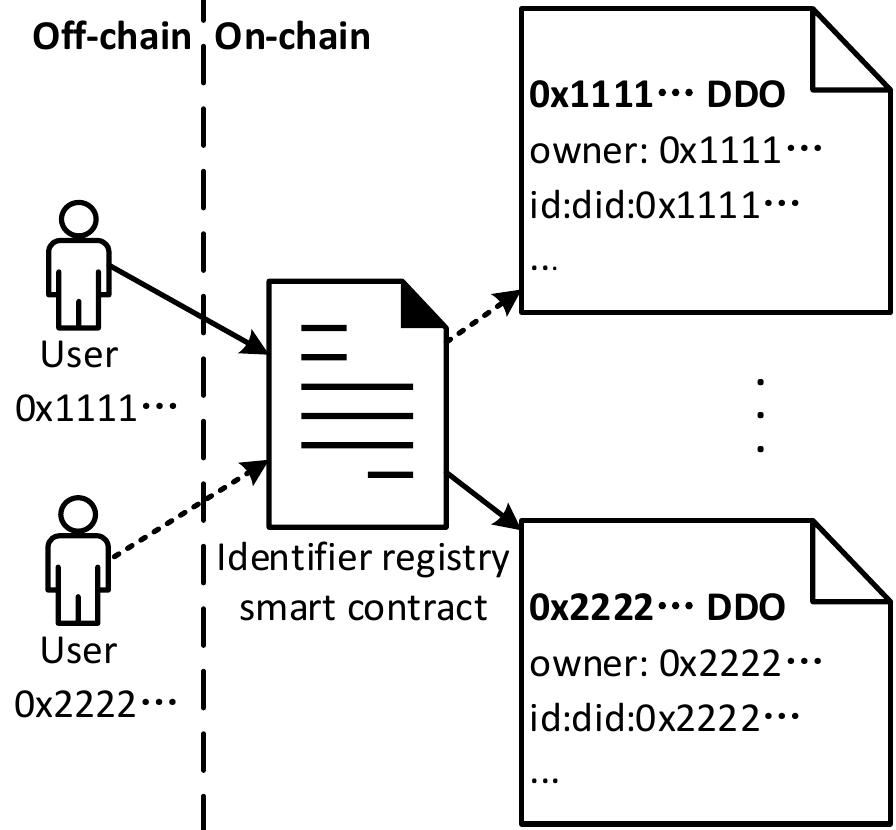}
	\caption{Dual Resolution Pattern}
	\label{dualPic}
\end{figure}

\vspace{0.5em}\noindent \textbf{Summary:} To establish interactions, two entities can mutually acquire each other's DDO to access information necessary for verification (e.g., public key) and communication (e.g., service endpoints for provided services). Fig.~\ref{dualPic} is a graphical representation of the pattern.

\vspace{0.5em}\noindent \textbf{Context:} 
In self-sovereign identity, entities interact with each other.

\vspace{0.5em}\noindent \textbf{Problem:}
When two or more entities want to establish interactions (e.g., for business purposes), each entity first needs to determine the target entity's basic information and ways of communicating before going further.


\vspace{0.5em}\noindent \textbf{Forces:} The problem requires 
balancing the following forces:
\begin{itemize}
  \item \textit{Interoperability.} For two parties to interact, the ways of communication have to be interoperable.
  \item \textit{Independence.} The interacting entities should remain independent of each other and one should not be able to pry on the other.
\end{itemize}

\vspace{0.5em}\noindent \textbf{Solution:} 
A DDO contains verification methods (i.e. public keys) and service endpoints (e.g., messaging service details) which can be utilised by an entity to establish interactions with the corresponding DID owner. Before any formal activity between two entities in a relationship, they should first \textit{mutually} resolve each other's DID and obtain the interaction information stored in DDO. Such a process is considered as ``Dual Resolution" and it forms the first step for any entity to establish an interoperation with its target entity.

\vspace{0.5em}\noindent \textbf{Consequences:} 

Benefits:
\begin{itemize}
  \item \textit{Interoperability.} The \emph{Dual Resolution} process allows the interacting entities in a relationship to obtain and check each other's basic information on verification and services provided for interoperability.
  
  \item \textit{Independence.} Each DDO stores necessary interaction information only for its corresponding DID, hence, an entity's different identities are independent from each other and cannot be correlated.
\end{itemize}

Drawbacks: 
\begin{itemize}
\item \textit{Privacy.} It is possible for an entity to store in DDO extra information about itself other than what is necessary for communication, such as social media accounts, personal websites. Doing so may increase the trustworthiness of the entity as more identity information is shared with others. However, this also increases the risk of unwillingly revealing the entity's personal information.
\end{itemize}

\vspace{0.5em}\noindent \textbf{Related patterns:}
\begin{itemize}
\item \textit{Identifier Registry} (Section~\ref{identifier registry}) Entities resolve each other's identifiers in \emph{Identifier Registry}.
\end{itemize}

\vspace{0.5em}\noindent \textbf{Known uses:}
The existing self-sovereign identity applications do not directly point out this feature as a provided functionality, but the users need to resolve each other's DID when there is potential interaction.

\subsubsection{Delegate List} \noindent \par
\label{recovery delegate registry}

\begin{figure}[!ht]
	\centering
	\includegraphics[width=0.71\columnwidth]{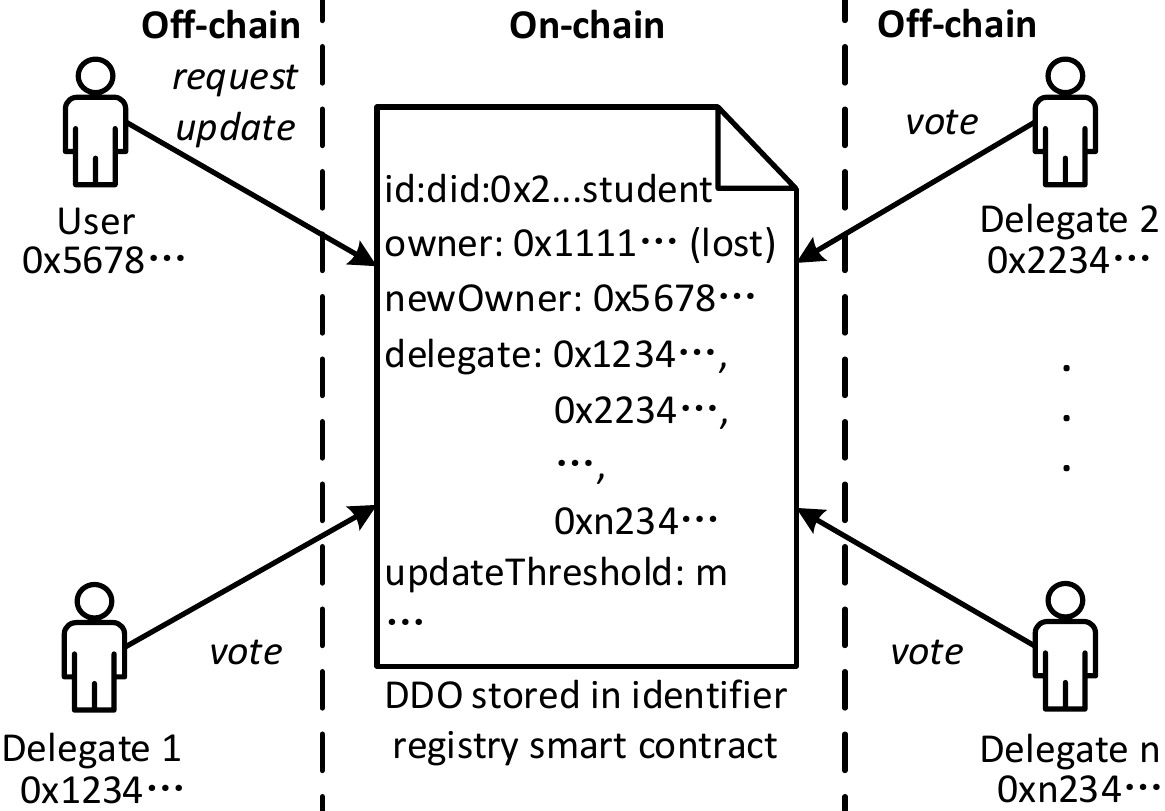}
	\caption{Delegate List Pattern}
	\label{updateByDelegatesPic}
\end{figure}

\vspace{0.5em}\noindent  \textbf{Summary:}
Each identifier maintains a list of delegates that can help the user recover the identity. Fig.~\ref{updateByDelegatesPic} is a graphical representation of the pattern.

\vspace{0.5em}\noindent \textbf{Context:} 
Each identity has a key pair to authenticate the transactions initiated by the user by means of digital signatures.

\vspace{0.5em}\noindent \textbf{Problem:} 
A master-key may be compromised/stolen by malicious hackers. A compromised master-key results in the loss of ownership over all sub-keys and corresponding identifiers. The hacker may utilise the identifiers to further steal the entity's identity data.

\vspace{0.5em}\noindent \textbf{Forces:} The problem requires balancing the following forces:
\begin{itemize}
  \item \textit{Compromised private key.} Authentication can be achieved by using digital signatures. Currently, many blockchain platforms do not provide a sound mechanism to recover compromised keys, thus key theft results in permanent loss of control over related identifiers.
  
  \item \textit{Non-reusability.} A compromised identity can be claimed of no more use, but its owner has to spent extra time, money, and energy to re-register a new identifier and rebuild all corresponding relationships.
\end{itemize}

\vspace{0.5em}\noindent \textbf{Solution:} 
\emph{Delegate List} relies on a web of trust architecture. This requires an identity owner to designate its own set of trustees that the owner trusts to assist in identity ownership update when the owner asks for it. An identifier maintains a list of recovery delegates and an update threshold that can help the user recover identity. These delegates can be individuals (such as family members or friends) or organisations (such as banks). If key loss happens, the original identity owner needs to request for ownership update using a new key pair, and a minimum number of the trustees (e.g. 2 out of 3) must sign a new identity record transaction respectively. When there are enough confirmations (i.e. reaching the threshold) of the new key pair, the ownership of the identifier is updated and thus the identity is recovered. A timelock period can be specified to prevent an attacker who tries to compromise an identity owner's key and immediately change the owner's identity records, including his/her designated trustees to prevent identifier ownership recovery.

\vspace{0.5em}\noindent \textbf{Consequences:} 

Benefits:
\begin{itemize}
  \item \textit{Tolerance of compromised private key.} This pattern can guarantee security by maintaining a group of delegates who can confirm a new key proposed by the identity owner to replace the compromised key, improving the tolerance of key compromise.
  
  \item \textit{Reusability.} The ownership of a compromised identity can be recovered and then put into use again, which mitigates the burden of rebuilding another same identity.
\end{itemize}

Drawbacks: 
\begin{itemize}
   \item \textit{Cost.} Recovering an identity requires setting up a delegate list in advance, and sending update requests/votes after the identity is compromised. Each operation sends a transaction to blockchain.
\end{itemize}

\vspace{0.5em}\noindent \textbf{Related patterns:} 
\begin{itemize}
\item \textit{Identifier Registry} (Section~\ref{identifier registry}) Delegates related to a specific identifier are stored in \textit{Identifier Registry}.

\item \textit{Hot \& Cold Wallet Storage (Section~\ref{hot-cold})} When being integrated into wallet applications, predefined delegates can replace key ownership if a key is compromised.

\item \textit{Key Shards} (Section~\ref{Key Shards}) \emph{Key Shards} can restore a lost key, while \emph{Delegate List} aims to replace a compromised key with a new one.

\item \textit{Multiple Authorisation}~\cite{xu2018pattern} \textit{Delegate List} is derived from \textit{Multiple Authorisation} that the change of a lost secret key in a DID requires enough confirmation.

\item \textit{Mutex Pattern}~\cite{wohrer2018smart} \textit{Mutex Pattern} can be applied to \textit{Delegate List}, ensuring that no any other operation can be conducted within an update procedure.
\end{itemize}

\vspace{0.5em}\noindent \textbf{Known uses:}
\begin{itemize}
  \item \emph{uPort}\textsuperscript{\ref{uport}} In uPort, the user's mobile device is the only place where stores the private key that controls a uPortID. To avoid the key loss issue caused by loss or theft of the mobile devices, users must nominate a group of delegates who can vote to replace the public key. Once a quorum is achieved by the delegates on the newly proposed public key, the lost public key is replaced with the new public key.
  
  \item \emph{Sovrin}\textsuperscript{\ref{sovrin}} Similar to uPort, Sovrin provides a key recovery mechanism for key recovery that relies on the user selecting a group of delegates. When the user requests the delegates to assist key recovery, a specified quorum of delegates must sign a new identity record transaction that validator nodes must verify.
  
  \item \emph{Baidu Cloud}\footnote{\url{https://cloud.baidu.com/}} Baidu implemented its SSI solution via Quorum, an Ethereum-based DLT. In each user's DDO, a ``recovery" attribute defines a list of public keys to recover the on-chain identity.
\end{itemize}

\subsection{Credential Design Patterns}

\subsubsection{Selective Content Generation} \noindent \par
\label{selective content}

\begin{figure}[!ht]
	\centering
	\includegraphics[width=0.54\columnwidth]{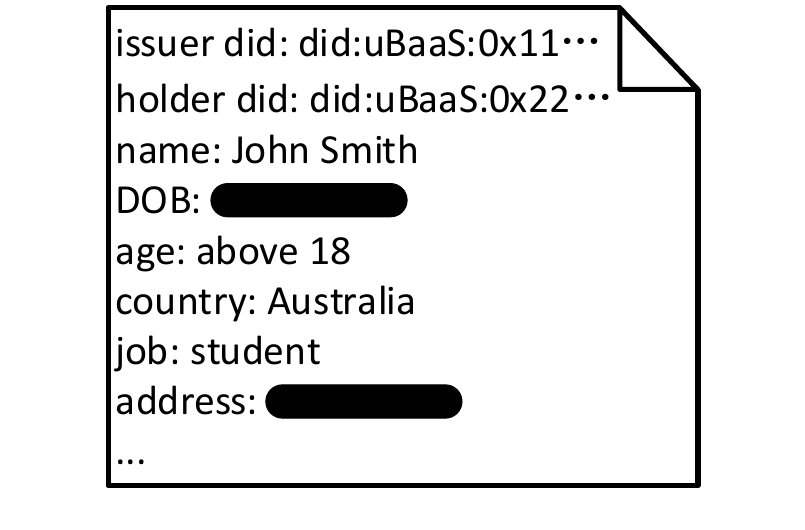}
	\caption{Selective Content Generation Pattern}
	\label{selectivePic}
\end{figure}

\vspace{0.5em}\noindent \textbf{Summary:} 
An issuer generates a customised credential according to a holder's specific requirements about credential contents. Fig.~\ref{credential} (a) is a graphical representation of the pattern.

\vspace{0.5em}\noindent \textbf{Context:} 
A verifier requires certain information to prove a holder's identity, thus, a holder only needs to share a credential with necessary data to the verifer.

\vspace{0.5em}\noindent \textbf{Problem:} 
If issuers publish general credentials to holders, a verifier can learn all identity data involved when only some particular attributes are needed. For instance, if a person shows his/her ID to identify the age, his/her address is presented either. This may cause data leak as extra information is provided.

\vspace{0.5em}\noindent \textbf{Forces:}
The problem requires balancing the following forces:
\begin{itemize}
  \item \textit{Privacy.} The disclosed credential should contain minimum amount of data necessary to identify some certain aspects of its holder.
  \item \textit{Specific requirements.} Each verifier may have specific requirements on inspecting a holder's identity facts. 
\end{itemize}

\vspace{0.5em}\noindent \textbf{Solution:}
\emph{Selective Content Generation} allows issuers to decide what identity attributes are contained in a credential. An issued credential needs to satisfy the target verifier's specific requirements of holder's identity, without revealing extra data. A credential with selective content disclosure can be generated via the following approaches\footnote{\url{https://w3c.github.io/vc-imp-guide/}}.

\begin{itemize}
  \item \textit{Atomic credentials.} 
  An issuer generates multiple credentials and each one only contains exactly one identity attribute about the holder. Consequently, the holder can flexibly disclose those required credentials to a verifier.
  
  \item \textit{Selective disclosure signatures.} 
  A general credential is issued to a holder, but some special signature schemes (e.g.  Camenisch-Lysyanskaya signatures) allow them to only reveal necessary information.
  
  \item \textit{Hashed values.} A general credential consists of multiple identity attributes, but each one is hashed with different nonce. When verifying a credential, a verifier can only validate those with the actual values provided by the holder, but cannot determine other hashed values.
  
  \item \textit{Zero-knowledge proof.} When proving certain identity attributes, a holder can protect its information by giving a range instead of precise value (e.g., age is over 18).
\end{itemize}

\vspace{0.5em}\noindent \textbf{Consequences:} 

Benefits:
\begin{itemize}
    \item \textit{Data minimisation.} A credential with selective content can disclose the identity data which satisfy the verifier's individual requests while keeping other identity data private.
\end{itemize}

Drawbacks: 
\begin{itemize}
   \item \textit{Cost.} Determining the identity data within a credential requires additional communication between holders and verifiers for learning the verification requirements, and between holders and issuers for discussing the credential content. Moreover, maintaining multiple credentials with different contents can incur extra costs.
\end{itemize}

\vspace{0.5em}\noindent \textbf{Related patterns:}
\begin{itemize}
\item \textit{Time-Constrained Access} (Section~\ref{accessible period}) \textit{Selective Content Generation} can work collaboratively with \textit{Time-Constrained Access}, to generate credentials with fine-grained specifications.

\item \textit{One-Off Access} (Section~\ref{one off}) \textit{Selective Content Generation} can also work collaboratively with \textit{One-Off Access}, to generate credentials with fine-grained specifications.

\item \textit{Blockchain Anchor} (Section~\ref{anchoring}) Off-chain credential contents need to be hashed and stored on-chain to preserve integrity.
\end{itemize}

\vspace{0.5em}\noindent \textbf{Known uses:}
\begin{itemize}
   \item \textit{uPort}\textsuperscript{\ref{uport}} A uPort user encrypts identity attributes using a symmetric encryption key before disclosure. The symmetric encryption key is then individually encrypted using a public encryption key owned by the other interacting party.
   \item \textit{Sovrin}\textsuperscript{\ref{sovrin}} A cryptographic technique known as a zero-knowledge proof is utilised in Sovrin. A verifier can check the authentication of an identity though the public key of the issuer, but never learns the actual data.
\end{itemize}

\subsubsection{Time-Constrained Access} \noindent \par
\label{accessible period}

\begin{figure}[!ht]
	\centering
	\includegraphics[width=0.535\columnwidth]{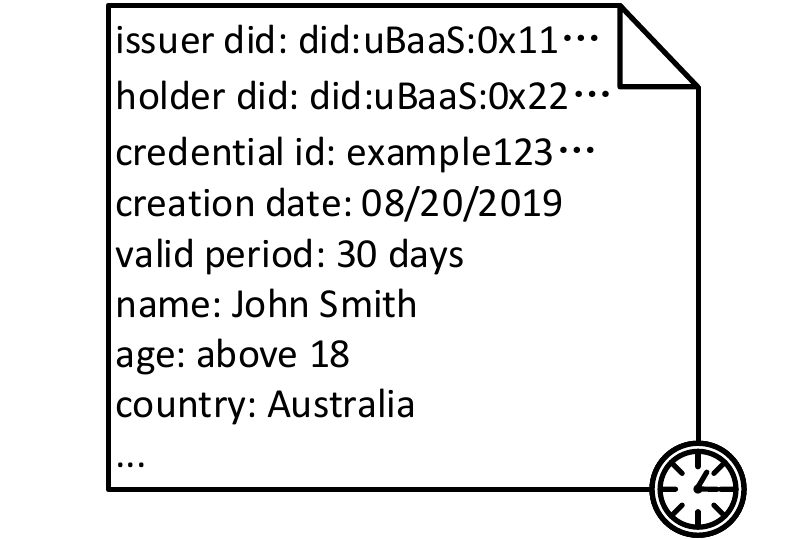}
	\caption{Time-Constrained Access Pattern}
	\label{TimePic}
\end{figure}

\vspace{0.5em}\noindent \textbf{Summary:} 
A holder can share a link which is redirected to the credential content only for a restricted accessible time period.
The verifier can only access the credential content within the determined time period. Fig.~\ref{credential} (b) is a graphical representation of the pattern.

\vspace{0.5em}\noindent \textbf{Context:} 
Usually an identification process lasts for a certain time period. After proving the identity of an entity, the presented credential has accomplished its mission and should not be accessed again.

\vspace{0.5em}\noindent \textbf{Problem:} 
After receiving a credential, a verifier then has the ability to access, read, and verify certain identity data of the holder. If the credential is long-term or even permanently effective, the verifier then can verify the credential after current identification process, which means that it can still access and check the holder's identity data when there is not a legitimate permission for proving the identity, resulting identity data leak.

\vspace{0.5em}\noindent \textbf{Forces:}
The problem requires balancing the following forces:
\begin{itemize}
  \item \textit{Privacy.} A holder's identity information should not be accessed or verified when the current identification process is finished.
  
  \item \textit{Inflexibility.} Verifiers have their own identification processes, which may take different amounts of time. The cost to generate and maintain credentials for the process inconsistency is huge.
\end{itemize}

\vspace{0.5em}\noindent \textbf{Solution:}
A holder is able to generate an identifiable link, and define its accessible period (e.g. certain days). The link can redirect to a page presenting credential content. Afterwards, the holder can share the time-constrained link to verifiers instead of the original credential itself. Within the predefined accessible period, a verifier can visit and verify the credential for identification without limit. Nevertheless, when the link is expired, there is no approach for the verifier to obtain credential content again.

\vspace{0.5em}\noindent \textbf{Consequences:} 

Benefits:
\begin{itemize}
  \item \textit{Privacy.} A holder can determine the accessible period of a shared link, which ensures that the holder's identity information can only be fetched within a particular identification process. An expired credential cannot be verified again. Consequently, a malicious verifier is unable to further utilised the identity data.
  
  \item \textit{Flexibility.} Shared links does not affect the original credential. Consequently, this pattern can be flexibly applied to a long-term effective credential, links with different accessible periods can be sent to different verifiers.
\end{itemize}

Drawbacks: 
\begin{itemize}
   \item \textit{Cost.} Storing the credential replicas requires more storage.
   
   \item \textit{Privacy.} A malicious verifier may take a photo of the credential when accessing it, then it has the credential content even if the shared link is expired. Although the compromised credential cannot reveal up-to-date information of the holder, the attacker still maintains historical identity attributes of the holder to some extent.
\end{itemize}

\vspace{0.5em}\noindent \textbf{Related patterns:}
\begin{itemize}
\item \textit{Selective Content Generation} (Section~\ref{selective content}) \textit{Time-Constrained Access} can work collaboratively with \textit{Selective Content Generation}, to generate credentials with fine-grained specifications.

\item \textit{One-Off Access} (Section~\ref{one off}) \textit{One-Off Access} can be seen as a derivative of \textit{Time-Constrained Access} under an extreme condition.
\end{itemize}

\vspace{0.5em}\noindent \textbf{Known uses:}
\begin{itemize}
    \item \textit{Snapchat}\footnote{\url{https://www.snapchat.com/}\label{Snapchat}} Snapchat is a social media in which users can share their photos and videos. Every shared photo or video is automatically deleted after a certain amount of time.
    
   \item \textit{Snappass}\footnote{\url{https://oneoffsecret.com/}\label{Snappass}} Snappass is a website where users can generate secret information and share it by URL. A user can set a valid time period to each secret, within which the secret information can be read.
\end{itemize}

\subsubsection{One-Off Access} \noindent \par
\label{one off}

\begin{figure}[!ht]
	\centering
	\includegraphics[width=0.54\columnwidth]{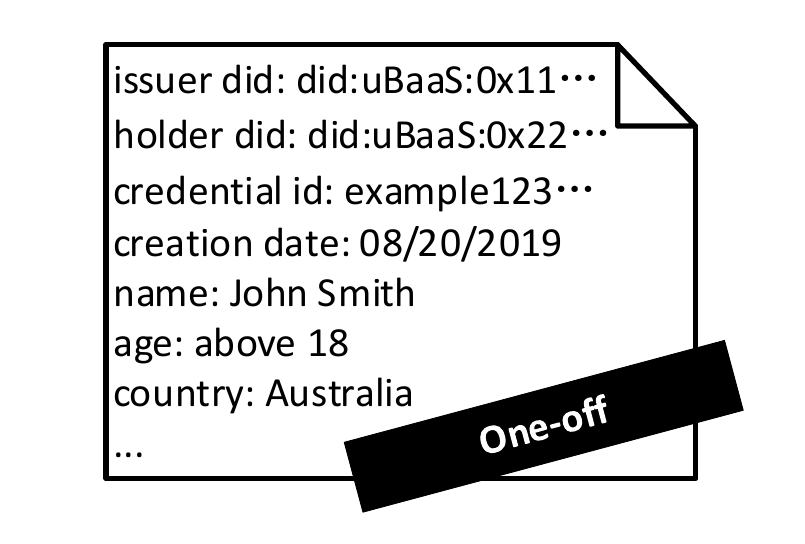}
	\caption{One-Off Access Pattern}
	\label{oneoffPic}
\end{figure}

\vspace{0.5em}\noindent \textbf{Summary:} 
A holder can share a ``one-off'' link which is redirected to the credential content one time only. This may be used to satisfy a temporary identification request. Fig.~\ref{credential} (c) is a graphical representation of the pattern.

\vspace{0.5em}\noindent \textbf{Context:} 
A verifier does not require a long-term effective credential but only needs to check the identity of a holder once for a specific purpose.

\vspace{0.5em}\noindent \textbf{Problem:} 
Sometimes an identification process does not require a strict verification procedure, but only needs to check the identity for once. For instance, travelling by train/airplane or going to a theme park only asks for checking credentials before entering. If a holder presents a long-term effective link redirecting to the credential content, a malicious verifier may access the holder's data illegally after identification process. This can be considered as an extreme version of \textit{Time-Constrained Access}.

\vspace{0.5em}\noindent \textbf{Forces:}
The problem requires balancing the following forces:
\begin{itemize}
  \item \textit{Privacy.} A holder's identity information should not be accessed or verified when the current identification process is finished.
  
  \item \textit{Inflexibility.} Using a long-term effective credential to satisfy a temporary credential request is not appropriate, as the holder cannot control the access to this credential after the temporary request.
\end{itemize}

\vspace{0.5em}\noindent \textbf{Solution:}
A holder is able to generate an identifiable link, which redirects to a one-off page presenting the credential content. One-off links can be shared with verifiers on some special occasions. After being visited once, the link becomes invalid that no one can use it to access the credential content.

\vspace{0.5em}\noindent \textbf{Consequences:} 

Benefits:
\begin{itemize}
  \item \textit{Privacy.} The identity attributes presented via one-off links can only be read and verified for once. Similar to \emph{Time-Constrained Access}, malicious verifiers cannot further violate holders' privacy via expired links. Information within the link becomes unauthentic as there is no approach to verify it.
  
  \item \textit{Flexibility.} \emph{One-Off Access} defines an extremely short period to visit the credential content, satisfying temporary identification requests.
\end{itemize}

Drawbacks: 
\begin{itemize}
   \item \textit{Privacy.} A malicious verifier may take a photo of the credential when accessing it, then it can read the content even if the one-off link is expired. Although information within an expired link can no longer be used to prove anything, this still can cause a privacy issue for the holder.
\end{itemize}

\vspace{0.5em}\noindent \textbf{Related patterns:}
\begin{itemize}
\item \textit{Selective Content Generation} (Section~\ref{selective content}) \textit{One-Off Access} can work collaboratively with \textit{Selective Content Generation}, to generate credentials with fine-grained specifications.

\item \textit{Time-Constrained Access} (Section~\ref{accessible period}) \textit{One-Off Access} can be seen as a derivative of \textit{Time-Constrained Access} under an extreme condition.
\end{itemize}

\vspace{0.5em}\noindent \textbf{Known uses:}
\begin{itemize}
    \item \textit{Snapchat}\textsuperscript{\ref{Snapchat}} Snapchat can automatically delete user-uploaded photos or videos when once read, according to the user's setting.

   \item \textit{Snappass}\textsuperscript{\ref{Snappass}} In Snappass, user-defined secret information are delete and cannot be recovered after being read for once.
\end{itemize}

\subsubsection{Blockchain Anchor} \noindent \par
\label{anchoring}

\begin{figure}[!ht]
	\centering
	\includegraphics[width=0.635\columnwidth]{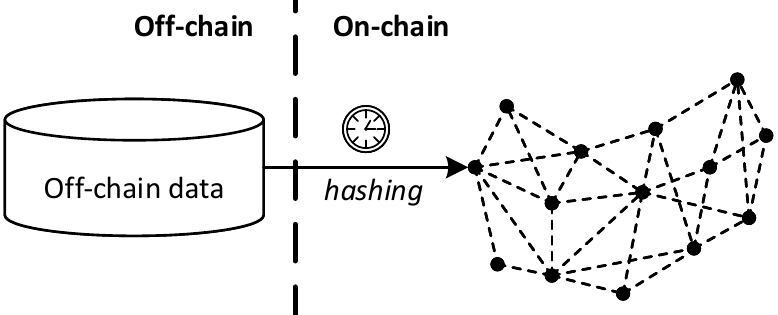}
	\caption{Blockchain Anchor Pattern}
	\label{anchoringPic}
\end{figure}

\vspace{0.5em}\noindent \textbf{Summary:} 
Instead of storing everything on-chain, one can periodically send the unique hash value of off-chain data to blockchain. Fig.~\ref{anchoringPic} is a graphical representation of the pattern.

\vspace{0.5em}\noindent \textbf{Context:} 
Blockchain's nature configurations may limit its performance when facing a large number of transactions.

\vspace{0.5em}\noindent \textbf{Problem:} 
Blockchain can ensure data integrity via storing data on-chain, but it costs real money to process the transaction in many public blockchain networks. In addition, according to the nature consensus mechanism, blockchain generates a block in a fixed period (i.e. block interval), which only includes a restricted number of transactions due to the block size. Consequently, blockchain's performance may be restricted when users frequently initiate transactions.

\vspace{0.5em}\noindent \textbf{Forces:}
The problem requires balancing the following forces:
\begin{itemize}
  \item \textit{Cost.}
  In a permissionless blockchain network, sending a transaction charges real money, thus, frequently storing data to blockchain is expensive. Even in a permissioned blockchain, each full node maintains a local replica of all historical transactions, handling plenty of transactions is also costly for physical storage.
  
  \item \textit{Scalability.}
  Blockchain has nature performance restriction that within a block interval, only a limited number of transactions can be processed while others need to wait in the transaction pool. Utilising blockchain to record every data change may result in the accumulation of waiting transactions in the pool.
\end{itemize}

\vspace{0.5em}\noindent \textbf{Solution:}
\emph{Blockchain Anchor} relies on the hashing technology that one does not need to store everything on-chain, but periodically sends the unique hash value of off-chain data to blockchain.

\vspace{0.5em}\noindent \textbf{Consequences:} 

Benefits:
\begin{itemize}
  \item \textit{Cost.}
  Anchoring reduces the cost of applying blockchain in terms of monetary payment and physical storage, as there are less transactions sent to blockchain.
  
  \item \textit{Scalability.} Anchoring keeps complex and tedious business processes off-chain, which enhances the scalability of blockchain-based systems that blockchain only needs to record the hash values of all kinds of data and files.
  
  \item \textit{Opacity.} Blockchain transactions are immutable and can be view by all participants. Thus, storing hash values rather than original information can preserve data privacy.
\end{itemize}

Drawbacks: 
\begin{itemize}
   \item \textit{Opacity.}
   Hash values are neither human-readable or can be restored into original files, which means that using anchoring may affect the transparency and auditability of on-chain data.
\end{itemize}

\vspace{0.5em}\noindent \textbf{Related patterns:}
\begin{itemize}
\item \textit{Selective Content Generation} (Section~\ref{selective content}) Off-chain credential contents need to be hashed and stored on-chain to preserve integrity.

\item \textit{Off-Chain Data Storage}~\cite{xu2018pattern} \textit{Blockchain Anchor} works similarly as \textit{Off-Chain Data Storage}, calculating the hash of raw off-chain and storing the hash values on-chain.

\item \textit{Low Contract Footprint Pattern}~\cite{eberhardt2017or} \textit{Blockchain Anchor} works similarly as \textit{Low Contract Footprint Pattern}, which concerns about the monetary cost and optimises the writing operations to blockchain.
\end{itemize}

\vspace{0.5em}\noindent \textbf{Known uses:}
\begin{itemize}
   \item \textit{Blockstack}\textsuperscript{\ref{blockstack}} Blockstack allows entities to register off-chain decentralised identifiers. To prove the existence of these off-chain identifiers, the system collects hashes of corresponding files and writes the hash values to blockchain.
    \item \textit{Chainpoint}\footnote{\url{https://chainpoint.org/}} Chainpoint is an open standard for creating a timestamp proof of any data, file, or process by generating a Merkle Tree, and publishing root of this tree to the Bitcoin blockchain.
    \item \textit{Laava} A blockchain-based system may need to serve different tenants simultaneously, who are not willing to expose private data to each other. Researchers from Data61 and Laava ID Pty. Ltd. designed a scalable platform architecture for multitenant blockchain-based systems \cite{ICSA2019}, in which every tenant has an individual permissioned blockchain to maintain their own data, while all tenant chains are anchored into a main chain periodically.
\end{itemize}
\section{Conclusion}
\label{conclusion}

We view blockchain as a fundamental component of large-scale decentralised software systems. In self-sovereign identity, blockchain provides an underlying computing infrastructure and decentralised pseudonym mechanism. In this paper, we summarise and propose 12 different design patterns associated with the lifecycles of three main objects (i.e. key, decentralised identifier, and credential) in blockchain-based self-sovereign identity. 
The pattern collection is considered as a guidance for architects to better design blockchain-based self-sovereign identity systems. 

\section*{References}


\end{document}